\documentclass[pra,twocolumn,showpacs,centertags]{revtex4}
\usepackage{amssymb,amsmath}
\usepackage{eufrak}
\usepackage{float}
\usepackage{epsfig}

\begin{document}
 \graphicspath{{.}} 
 
 \title{Mean-field phase diagram for Bose-Hubbard Hamiltonians with random
     hopping}
\author{\firstname{Pierfrancesco} \surname{Buonsante}}
\affiliation{Dipartimento di Fisica, Torino Politecnico,
             Corso Duca degli Abruzzi 24, I-10129 Torino, Italy}

\author{\firstname{Francesco} \surname{Massel}}
\affiliation{Dipartimento di Fisica, Torino Politecnico,
             Corso Duca degli Abruzzi 24, I-10129 Torino, Italy}

\author{\firstname{Vittorio} \surname{Penna}}
\email[corresponding author:]{vittorio.penna@polito.it}
\affiliation{Dipartimento di Fisica, Torino Politecnico,
             Corso Duca degli Abruzzi 24, I-10129 Torino, Italy}

\author{\firstname{Alessandro} \surname{Vezzani}}
\affiliation{Dipartimento di Fisica, Universit\`a  degli Studi di Parma and
             C.N.R.-I.N.F.M., 
             Parco Area delle Scienze 7/a, I-43100 Parma, Italy}

\begin{abstract}
  The zero-temperature phase diagram for ultracold Bosons in a random 1D
  potential is obtained through a site-decoupling mean-field scheme performed
  over a Bose-Hubbard (BH) Hamiltonian whose hopping term is considered as a
  random variable. As for the model with random on-site potential, the
  presence of disorder leads to the appearance of a Bose-glass phase. The
  different phases --i.e. Mott insulator, superfluid, Bose-glass-- are
  characterized in terms of condensate fraction and superfluid fraction.
  Furthermore, the boundary of the Mott lobes are related to an off-diagonal
  Anderson model featuring the same disorder distribution as the original BH
  Hamiltonian.
\end{abstract}
\maketitle

\section{Introduction}
   \label{sec:intro}
In the framework of ultracold atom physics, the experimental tunability of
control parameters pertaining each model Hamiltonian has provided a powerful
tool to investigate situations of fundamental physical interest 
\cite{A:Jaksch}. 

One of the most intriguing features about ultracold atoms is the possibility
to engineer a defect-free periodic potential, as opposed, for instance, to the
typical framework of solid-state physics. However, on the one hand the
interplay between disorder and interactions in Bosonic systems has attracted
much theoretical attention since the seminal work by Fisher \cite{A:Fisher},
and on the other several techniques such as laser speckle field \cite{A:Lye},
the superposition of different optical lattices with incommensurate lattice
constants \cite{A:Roth03,A:Damski2003,A:Fallani}, have proven the
experimental relevance of disordered systems of ultracold atoms.

In the present paper we will deal with the effect of disorder on the
zero-temperature phase diagram of bosonic atoms loaded onto a 1D lattice whose
properties can be described in terms of Bose-Hubbard Hamiltonian. In
particular, we will focus on the case where the disorder affects the hopping
term
\begin{eqnarray}
  \label{eq:BHdis}
   H  =&&  \sum_{m=1}^M \frac{U}{2} a_m^\dag a_m^\dag a_m a_m 
          -\mu\, a_m^\dag a_m \nonumber \\  
       && -\sum_{m,m'} J_{m,m'} a_m^\dag  a_{m'} + h.c. 
\end{eqnarray}
where $a_m^\dag$ ($a_m$) represents the creation (destruction) operator on
site $m$. The Hamiltonian parameters $U$, $J_{m,m'}$ represent the two-body
interaction and the (random) hopping amplitude between neighboring sites
respectively.

In the recent past, many authors have approached the analysis of the disordered
BH model with various techniques such as field-theoretic approaches
\cite{A:Fisher,A:Wallin,A:Svistunov1996,A:Pazmandi},
decoupling (or Gunzwiller) mean-field approximations
\cite{A:Damski2003,A:Sheshadri1993,A:Krauth1992,A:Sheshadri1995,A:Krutitsky},
quantum Monte-Carlo simulations
\cite{A:Scalettar1991,A:Lee2004},
among many others, e.g.
\cite{A:Singh,A:Freericks1996,A:Pai,A:Rapsch,A:Pugatch}.

Following \cite{A:Buonsante06CM}, we employ a site-decoupling mean-field
approximation (SDMFA), which allows to capture all of the essential features
of the phase diagram of the model \eqref{eq:BHdis}. The phases of the
zero-temperature phase diagram are determined through the calculation of two
different observables: the condensate fraction, defined as the largest
eigenvalue of the one-body density matrix and the superfluid fraction, defined
as the system response to the coupling to an external field \cite{A:Roth03,A:Penrose}.

At zero temperature, as already pointed out in \cite{A:Fisher} for the on-site
disordered BH model, we expect the presence of three phases: the
Mott-Insulating (MI) phase, where both condensate fraction and superfluid
fraction are zero, the superfluid phase, where both superfluid phase and
condensate fraction are different from zero, and, finally, the Bose-Glass
phase, which has zero superfluid fraction and finite condensate fraction,
which represents a typical feature of disordered hopping systems.

In Section \ref{sec:MF}, we introduce the site decoupling mean-field
approximation for the case given by Hamiltonian \eqref{eq:BHdis} and we depict
the zero-temperature phase diagram, discussing similarities and differences
between the random-hopping and the random on-site potential case.
 
In Section \ref{sec:stab}, we discuss the stability of the Mott phase through
the stability analysis of the recurrence map induced by the SDMFA. This
analysis will be performed comparing the results obtained by numerical exact
diagonalization and analytical results based on random-matrix theory.

\section{Site-decoupling mean-field scheme}
\label{sec:MF}

The SDMFA was introduced in Ref.  \cite{A:Sheshadri1993}, and relies on the
approximation
\begin{equation}
\label{E:mfa}
a_m^\dag a_{m'} = a_{m'}^\dag \alpha_m + a_m \alpha_{m'}^* - \alpha_m \alpha_{m'}^*
\end{equation}
where the $\alpha_m$'s are mean-field variables to be determined
self-consistently.  The above posit turns the BH Hamiltonian \eqref{eq:BHdis}
into a mean-field Hamiltonian that is the sum of on-site contributions.
\begin{eqnarray}
{\cal H} & = & \sum_m {\cal H}_m + J \sum_{m \, m'} \alpha_m^* A_{m \, m'} \alpha_m'
\label{E:MFH} \\
{\cal H}_m &=& \frac{U}{2}n_m\left(n_m-1\right) -\mu\, n_m \nonumber\\
&-& J (\gamma_m a_m^\dag + \gamma_m^* a_m) 
\label{E:sMFH}
\end{eqnarray}
where $n_m=a_m^\dag a_m$ is the usual bosonic number operator and the disorder
related to the hopping term has been embedded into the adjacency matrix
$A_{m,m'}$. For nearest-neighbor hopping on a 1D system the adjacency matrix
can be written as
\begin{equation}
 \label{eq:adjMord2}
 A_{m',m}= e^{i\,\phi}s_{m-1}\delta_{m',m-1} + e^{-i\,\phi}s_m\delta_{m',m+1}
\end{equation}
where $\phi$ takes into account a possible coupling to an external field
(\textit{Peierls factors}), $s_m \in \mathbb{R}$ accounts for a possible
inhomogeneity of the hopping amplitude. Here we consider a 1D lattice
comprising $N$ sites with periodic boundary conditions, i.e.
\begin{equation}
  \label{eq:pbc}
  s_{N+1}=s_1, \quad s_0=s_{N}.
\end{equation}

The ground state of Hamiltonian (\ref{E:MFH}) is
clearly a product of on-site states,
\begin{equation}
\label{E:MFgs}
| \Psi \rangle = \bigotimes_m | \psi_m \rangle
\end{equation}
and the requirement that the relevant energy is minimized results in the
self-consistency constraint \cite{B:Sachdev}

\begin{equation}
\label{E:SCc}
\gamma_m =  \sum_{m'} A_{m \, m'} \alpha_m, \qquad \alpha_m = \langle \psi_{m}| a_{m} |\psi_{m}\rangle
\end{equation}
The decoupling mean-field approach has proved to provide satisfactory
qualitative phase diagrams for homogeneous lattices
\cite{A:Sheshadri1993,A:vanOosten,A:Ferreira,A:Bru,A:Buonsante04} and superlattices
\cite{A:Buonsante04,A:Rey03}. 

\subsection{Numerical Simulation}
\label{sec:num}
In the present section we report the results of the application of SDMFA to
the zero-temperature phase diagram calculation.

\begin{figure}
 \label{fig:hop}
   \epsfig{figure=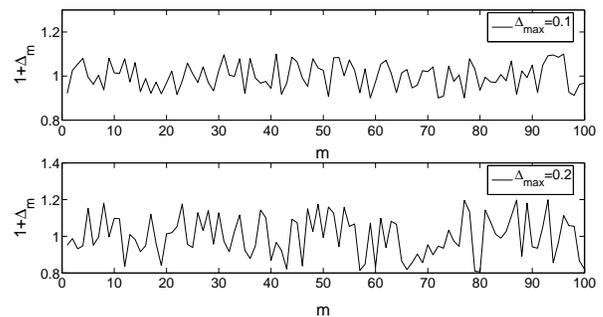,width=0.5\textwidth}
\caption{Values of $\Delta_m$ taken into account for the determination of the
  phase diagrams. Upper plot has weaker disorder amplitude ($\Delta_{\max}=0.1$)
  compared to the lower one  ($\Delta_{\max}=0.2$).}
\end{figure}

For our calculations we have considered a lattice formed by 100 sites, with
values of the adjacency matrix given by
\begin{equation}
  \label{eq:hop}
  A_{m,m'}= (1+\Delta_m)\delta_{m',m-1}+(1+\Delta_{m+1})\delta_{m',m+1} 
\end{equation}
where $\Delta_m$ is an uncorrelated random variable uniformly distributed
between $\Delta_{\max}$ and $-\Delta_{\max}$, with the
constraint $\Delta_{\max}<1$ in order to preserve $J_{m,m'}>0$. In Fig.
\ref{fig:hop}, we have represented two explicit realizations of disorder taken
into account for subsequent mean-field calculations of phase diagrams.

The value of $\Delta_{\max}$ has been kept small enough to ensure reasonable
self-averaging for the system under investigation. Higher disorder amplitudes
would require larger chains in order to obtain results independent of the
specific disorder realization.

The different phases have been identified through the evaluation of two
observables, namely the \textit{condensate fraction}, defined as the largest
eigenvalue of the one-body density matrix
\begin{equation}
  \label{eq:condfrac}
   f_c=max(\rho_{m,m'}),  \qquad \rho_{m,m'}=\langle a^\dagger_m a_m' \rangle/N,
\end{equation}
and the \textit{superfluid fraction} defined as
\begin{equation}
  \label{eq:sffrac}
  f_s=\lim_{\phi \rightarrow 0} \frac{E(\phi)-E(0)}{J \langle N \rangle\phi^2}.
\end{equation}
where $E(\phi$ is the ground-state energy corresponding to the presence of a
Peierls phase $\phi$ in the hopping term \eqref{eq:adjMord2}. 

The MI phase is characterized  by $f_c=f_s=0$,
the superfluid phase (SF) by $f_c \neq 0$ and $f_s \neq 0$, while the phase where
$f_c\neq 0$ and $f_s = 0$ is recognized as the Bose-glass (BG) phase \cite{A:Fisher}. 

In absence of disorder, the variation of the control parameters -- chemical
potential and hopping amplitude-- always leads to a transition from a phase
with both $f_s$ and $f_c$ equal to zero to a phase with $f_s$ and $f_c$
different from zero, excluding then the presence of a Bose-glass phase.

\begin{figure}[t!]
 \label{fig:lobe}
  \epsfig{figure=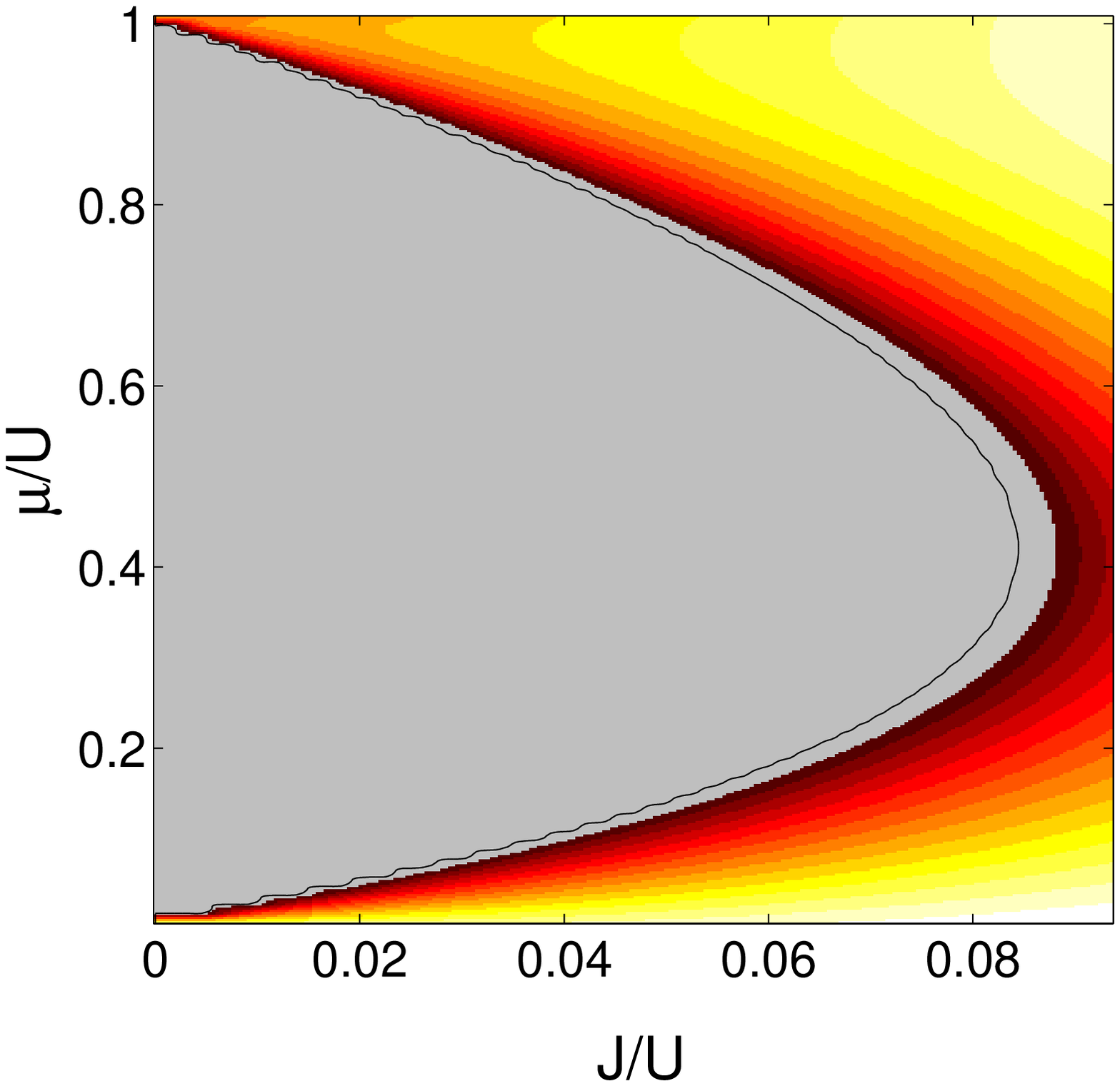,width=0.5\textwidth}
  \epsfig{figure=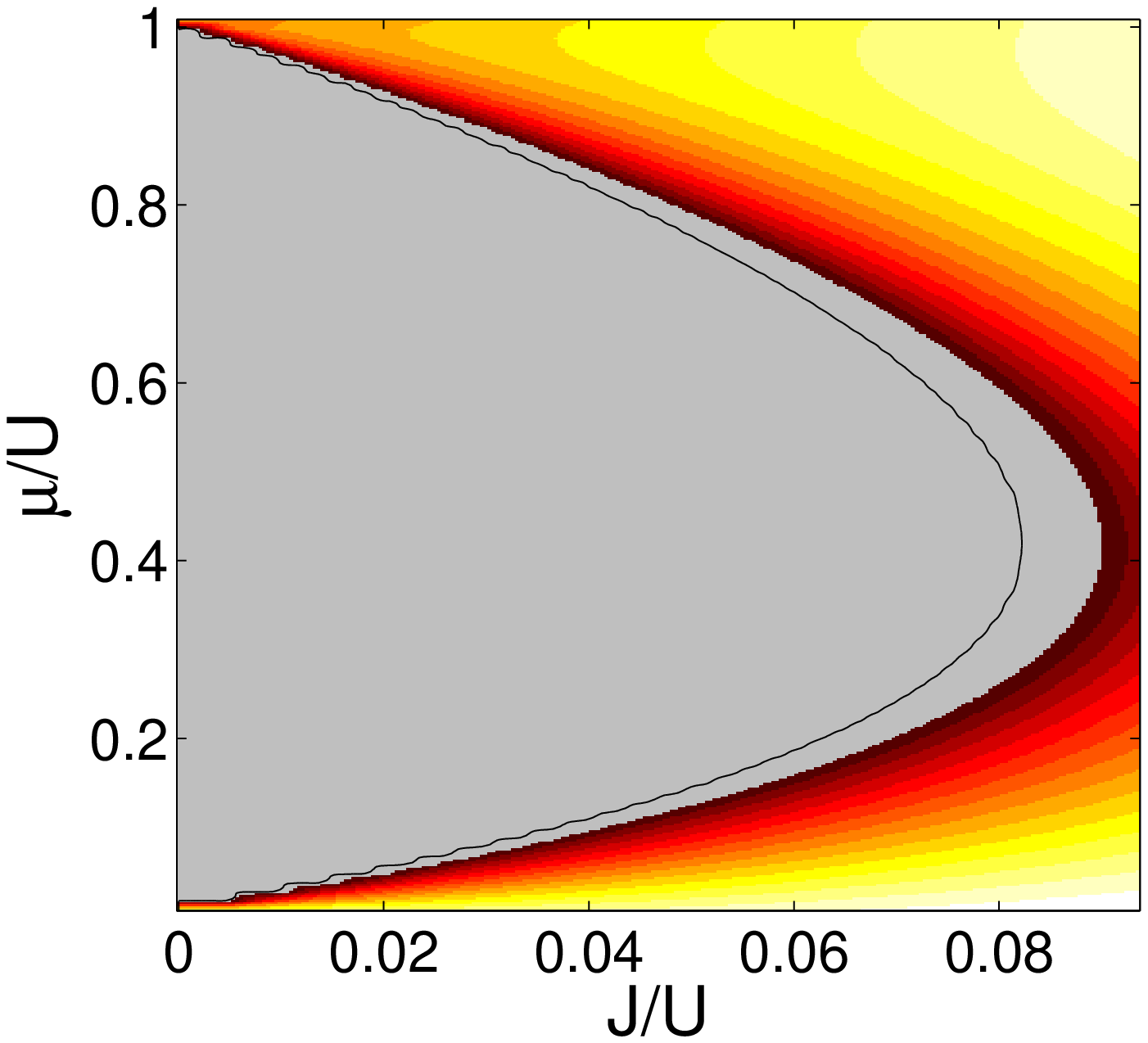,width=0.5\textwidth}
\caption{Plot of the zero-temperature phase diagram for
  $\Delta_{\max}=0.1$(upper panel) and $\Delta_{\max}=0.2$ (lower panel). In
  black we have plotted the border between the MI and the BG phase, while the
  colored background corresponds to the superfluid fraction (grey=$0$). It is
  possible to notice how the BG phase covers an increasing area as
  $\Delta_{\max}$ is increased.}
\end{figure}

On the other hand, if on-site disorder is present, the MI phase is (possibly)
separated from the SF phase by a BG phase. Likewise, when disorder affects the
hopping term alone a BG crops up, as it is shown in Fig. \ref{fig:lobe}.
However the distribution of the BG phase in the parameter space is
qualitatively different. For example, with on-site disorder MI and SF phase
are separated by a BG phase as $J$ goes to zero, while with disorder on the
hopping term the BG phase tends to disappear for small $J$
\cite{A:Buonsante06CM}. This region of the phase diagram seems to be a good
starting point for future investigations by means of a cluster MF approach
\cite{A:loophole,A:LoopLP}, because it would reveal possible MI phases
that are not detectable through the single-site mean-filed technique
implemented in the present paper.

In Fig. \ref{fig:lobe} we have represented the first lobe of the zero
temperature phase diagram as obtained by SDMFA for $\Delta_{\max}=0.1$ and
$\Delta_{\max}=0.2$. 

\section{Stability of the Mott phase}
\label{sec:stab}
In the present Section we introduce a procedure to determine the border of the
MI phase which is based on the stability of the self-consistency map induced
by the mean-field procedure.

As a first consideration, it is possible to state that the condition $\gamma_k
= 0$ for every $k$ corresponds to the gapped insulating phase of the
mean-field Hamiltonian \ref{E:MFH}.  Indeed, in this case the local
ground-states \eqref{E:MFgs} are eigenvectors of the local number operator
$a^\dag a$
\begin{equation}
\label{E:gsAc2}
|\psi \rangle = |n\rangle = \frac{(a^\dag)^n}{n!}|0\rangle,\quad n = \lceil \frac{\mu}{U} \rceil,
\end{equation}
where $\lceil x \rceil$ denotes the smallest integer larger than $x$, and
hence the mean-field ground-state (\ref{E:MFgs}) is a pure Fock state.  As for
the ordered Bose-Hubbard Hamiltonian, the relevant on-site energy is
\begin{equation}
\label{E:lgsE}
\epsilon_n = \frac{U}{2} n(n-1) + \mu n  
\end{equation}
Hence $\langle \psi|a|\psi \rangle = 0$ at every site, and the
self-consistency constraint \eqref{E:SCc} is satisfied. In other terms
$\gamma_k = 0$ is a fixed point of the map defined by Eq. \eqref{E:SCc}. The
gapped insulating phase is stable as long as this fixed point is stable.  This
is true as long as the magnitude of the maximal eigenvalue of the matrix
$\Lambda$ appearing in the linearized version of Eq.\eqref{E:SCc} is smaller
than 1, see \eqref{eq:linSelfC}. In order to determine $\Lambda$, we assume
$|\gamma_k| \ll 1$ and
consider the (mean-field) kinetic term in Hamiltonian \eqref{E:sMFH} as 
perturbative.  
If first order perturbation theory is performed one gets
\begin{equation}
\label{eq:linSelfC}
\langle a_m \rangle = \frac{J}{U} F\left(\frac{\mu}{U}\right) \sum_{m'} A_{m\,m'} \langle a_{m'} \rangle
\end{equation}
where
\begin{equation}
F\left(x\right) = \frac{x+1}{(\lceil x\rceil-x)(x-\lfloor x\rfloor)}
\end{equation}
Hence the linearized version of the self-consistency map Eq. \eqref{E:SCc} can be written as
\begin{equation}
\langle a_m \rangle = \frac{J}{U} \sum_{m'} \Lambda_{m\,m'} \langle a_{m'} \rangle 
\end{equation}
where the matrix $\Lambda$ is proportional to the adjacency matrix $A$:
\begin{equation}
\Lambda =  F\left(\frac{\mu}{U}\right)  A.
\end{equation}
Recalling the criteria for the stability of linear maps, the fixed point
$\langle a_m \rangle = 0$ (equivalent to $\gamma_m = 0$) is stable whenever
\begin{equation}
\label{eq:StabCond}
\frac{J}{U}\leq \frac{1}{|\tilde{\lambda}_{\max}|}
\end{equation}
where $\tilde{\lambda}_{\max}$ the eigenvalue of $\Lambda$ with the largest
magnitude. 

The problem of determining the maximal eigenvalue of the matrix $\Lambda$ can
be reduced to the calculation of the maximal eigenvalue of the ( tridiagonal )
adjacency matrix $A$. Note that $A$ is basically the one-particle Hamiltonian
for a non-interacting off-diagonal Anderson model (random hopping model). If
$\lambda_{\max}$ is the maximal eigenvalue of the adjacency matrix, Eq.
\eqref{eq:StabCond} can be recast as

\begin{equation}
  \label{eq:StabCond2}
  \frac{J}{U}\leq \frac{1}{|\lambda_{\max}|F\left(\frac{\mu}{U}\right)}.
\end{equation}

We have dealt with the calculation of the eigenvalues of $A$ both numerically
and analytically. The analytical approach consists in the determination of
the spectral density of the matrix, given the probability distribution of the
elements of the matrix following the approach outlined by Dyson for and
harmonic-oscillator chain \cite{A:Dyson}, while the former simply consists in the direct
numerical evaluation of the matrix eigenvalues.

\subsection{Numerical analysis}
\label{sec:NumAn}
\begin{figure}[t!]
  \centering
  \epsfig{figure=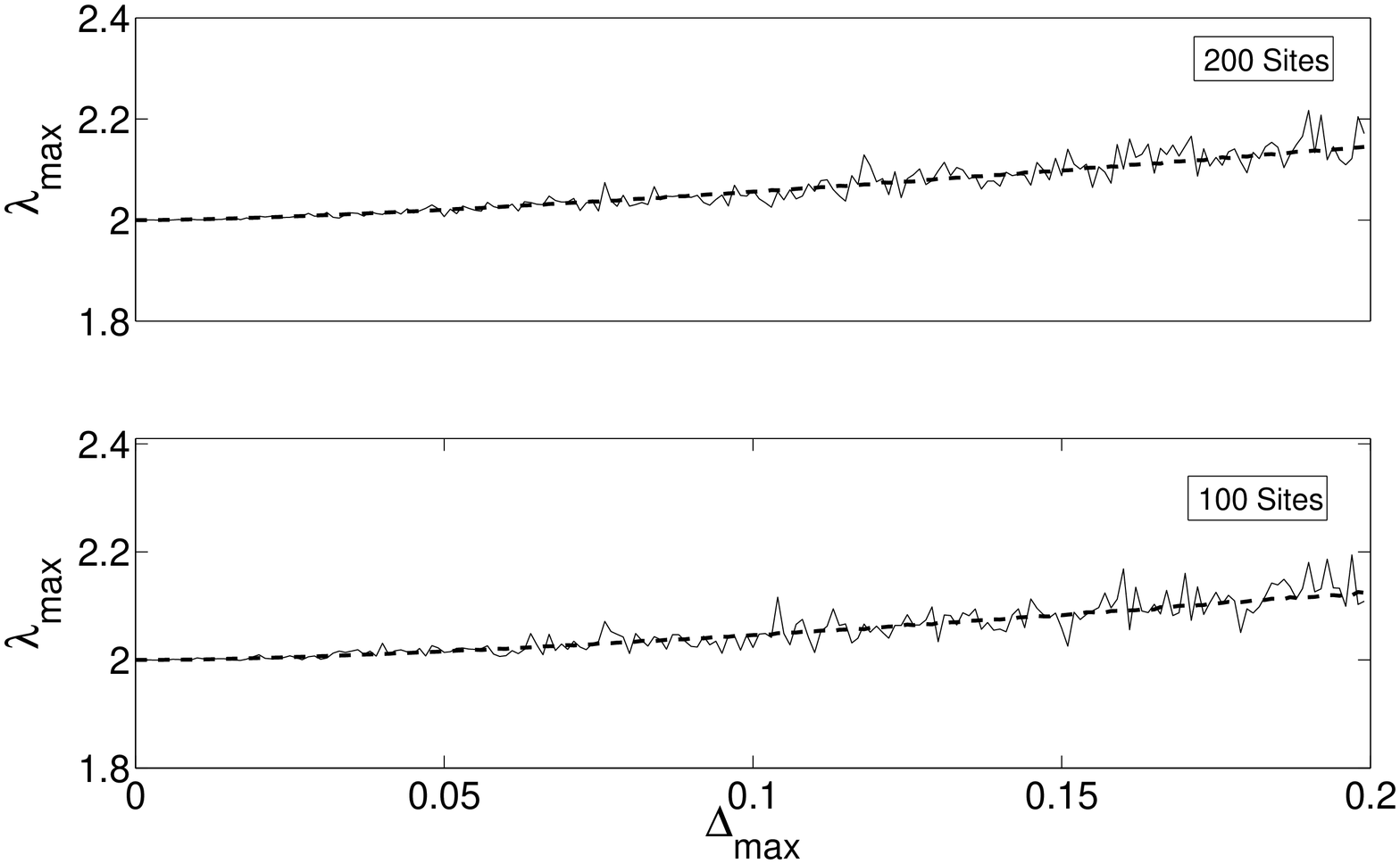,width=0.5\textwidth}
   \caption{$\Delta_{\max}$-dependence of the largest eigenvalue of the
     adjacency matrix $A$, for 200 sites (upper panel) and 100 sites (lower
     panel). By comparison between the single realization plots (full lines)
     and the averaged ones (dashed lines),it is possible to see how,
     increasing the lattice size, the effect of the specific randomness
     realization is decreased}
  \label{fig:lmax}
\end{figure}

In Fig. \ref{fig:lmax} we report the maximum eigenvalue $\lambda_{\max}$ for
the random adjacency matrix $A$ as a function of the strength of disorder for
both a single realization and a disorder average over 1000 samples.  The two
panels refer to different lattice sizes. In every case, it is evident that the
maximal eigenvalue $\lambda_{\max}$ grows with increasing disorder strength,
$\Delta_{\max}$ .

The stability condition given by Eq. \eqref{eq:StabCond2}, along with the
above considerations about the $\lambda_{\max}$-dependence from
$\Delta_{\max}$ are in agreement with the considerations of Section
\ref{sec:num} , as far as the MI phase is concerned. In particular
$\lambda_{\max}$ can be thought of as a shrinking factor for the MI lobe when
compared to a homogeneous situation.
\begin{figure}[t!]
  \centering
  \epsfig{figure=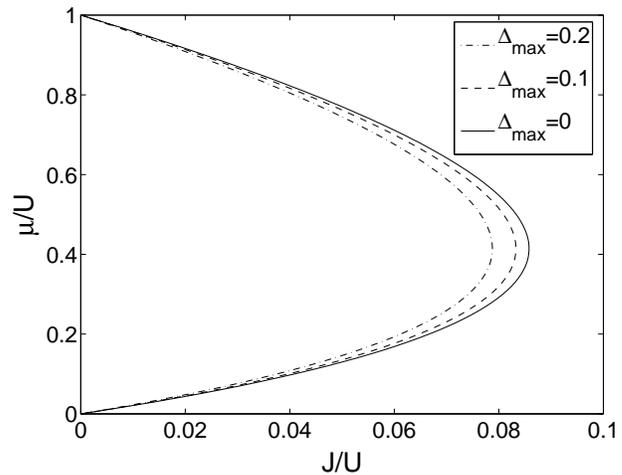,width=0.5\textwidth}
  \caption{Boundary of the MI region for different values of $\Delta_{\max}$.}
  \label{fig:lobStab}
\end{figure}

\subsection{Analytical solution}
\label{sec:An}

In this section we would like to describe an analytical method to obtain the
largest eigenvalue of a (possibly infinite) random adjacency matrix whose
entries are given by Eq. \eqref{eq:hop}. For some particular disorder
distributions it is possible to carry through the analytical calculation and,
as a consequence, solve Eq. \eqref{eq:StabCond2} without finite-size
effects.

It is worth mentioning that, in principle, the information gained through this
approach is richer. In fact for some specific realizations of disorder it is
possible to obtain the full integrated density of states $M(z)$ and not simply
the largest eigenvalue of the adjacency matrix.
  
The solution of the problem can be obtained following the approach proposed by
Dyson for the solution of a linear chain of harmonic oscillators.

Here we will sketch Dyson's approach, outlining the connection with our
problem, providing the portion of $M(z)$ needed to obtain the largest
eigenvalue in view of Eq.
\eqref{eq:StabCond2}.

In \cite{A:Dyson}, the problem of a linear harmonic chain with springs of
random elastic is reformulated as a tridiagonal matrix diagonalization
problem. The matrix $\Lambda$ has the following form
\begin{equation}
  \label{eq:SpM}
  \Lambda_{J+1,J}=-\Lambda_{J,J+1}=i\lambda_J^{1/2}
\end{equation}
which is related to the matrix $A$ defined in Eq. \eqref{eq:adjMord2} by a
unitary transformation $U(\theta)$
\begin{equation}
  \label{eq:U}
  U_{J,K}(\theta)=\delta_{J,K}\exp(i \theta J)
\end{equation}
with
\begin{equation}
  \label{eq:LUAU}
   \Lambda=U(\theta)\,A\,U(\theta)
\end{equation}
and 
$$
\theta=\frac{\pi}{2}-\phi 
$$
and setting $s_J=\lambda_J^{1/2}$ in Eq. \eqref{eq:adjMord2}.  The
unitarity of $U(\theta)$ allows to state that the eigenvalues of $\Lambda$ are
equal to the eigenvalues of $A$, hence the procedure followed in
\cite{A:Dyson} can be directly mapped onto our problem.

The core of this approach resides in the definition of the characteristic
function of the chain
\begin{equation}
  \label{eq:Om}
  \Omega(x)=\lim_{N\rightarrow \infty}\frac{\sum_j \log(1+x\omega_j^2)}{N}
\end{equation}
where $N$ is the size of the matrix, and $\omega_j$ the desired eigenvalues of
the matrix under consideration.

The density of states $D(z)$ and the integrated density of states 
$$
M(z)=\int_0^z D(z') dz'
$$
can be obtained from the characteristic function
through the following relation
\begin{equation}
  \label{eq:D}
  D(z)=-\frac{1}{z^2 \pi}{\rm Im}\left[\lim_{\epsilon \rightarrow 0 }  \Omega'(-\frac{1}{z}+i\epsilon)\right]
\end{equation}
having defined 
$$
\Omega'(x)=\frac{d \Omega}{d x}.
$$

The determination of $\Omega(x)$  is obtained through a power series
expansion  
\begin{equation}
  \label{eq:OmPow}
  \Omega(x)=\lim_{N \rightarrow \infty} \frac{1}{N} \sum_{n=1}^\infty (-1)^{n-1} Tr(\Lambda^{2n}).  
\end{equation}
The determination of $Tr(\Lambda^{2n})$ leads to the following relation
\begin{equation}
  \label{eq:OmXi}
   \Omega(x)=\lim_{N \rightarrow \infty} \frac{1}{N} \sum_{a=1}^{N}
   \log\left[1+\xi(a) \right] 
\end{equation}
having defined $\xi(a)$ through the continued fraction
\begin{equation}
  \label{eq:Xi}
  \xi(a)=x\lambda_a/\left(1+x\lambda_{a+1}/\left(1+x\lambda_{a+2}/\left(\dots\,
  \right.\right.\right..
\end{equation}
If, as it may be safely assumed in our case, the various values of $\lambda_a$
are independent random variables with probability distribution $G(\lambda)$,
the variable 
$$
\xi(a)=\frac{x\lambda_a}{1+\xi(a+1)}
$$ 
will have probability distribution $F(\xi)$ satisfying the following integral
equation
\begin{equation}
  \label{eq:Fint}
  F(\xi)=\int_0^\infty F(\xi') G\left[\xi(1+\xi'/x)\right]\frac{1+\xi'}{x} d\xi'.
\end{equation}
With the normalization condition
\begin{equation}
  \label{eq:NormF}
  \int_0^\infty F(\xi)d\xi=1
\end{equation}
we obtain 
\begin{equation}
  \label{eq:OmUnc}
  \Omega(x)=2\int_0^\infty F(\xi)\log(1+\xi)d\xi. 
\end{equation}

If we assume a Poissonian form for $G(\lambda)$ 
\begin{equation}
  \label{eq:Gpois}
  G_n(\lambda)=\frac{n^n}{(n-1)!}\lambda^{n-1}\exp(-n\lambda),
\end{equation}
Eq. \eqref{eq:Fint} has an analytical solution. Hence it is possible to obtain 
the integrated density of states in closed form in terms of integral functions. 

The integrated density of states for
$A$ can be simply obtained back by posing
\begin{equation}
  \label{eq:Ma}
  M^A(z)=M(z^2)
\end{equation}
since in \cite{A:Dyson} $M(z)$ is defined as the proportion of eigenvalues for
which $\omega_j^2<z$, while $M^A(z)$ as the proportion of eigenvalues with
$\lvert \omega_j\rvert<z$. As a simple check, we provide here the expression for
$M^A_\infty(z)$, corresponding to the case without disorder, i.e. $n=\infty$
in Eq. \eqref{eq:Gpois},
\begin{equation}
  \label{eq:Minf}
  \begin{array}{lll}
    M^A_\infty(z)&=\frac{1}{\pi}\arccos\left[1-1/2 z^2\right] & z<2 \\
                 &=1                                       & z>2.
  \end{array}
\end{equation}
which, as expected, coincides with the well known result for a 1D homogeneous system.
 
On the other hand, if we are interested in the determination of the maximal
eigenvalue of $A$ in presence of (weak) disorder whose distribution can be
related to that expressed by Eq. \eqref{eq:Gpois}, we can consider a large-$n$
expansion of $M^A_n(z)$ for $z>2$.
The expression of $M^A_n$ in this case is given by
\begin{equation}
  \label{eq:LargeM}
  M^A_n(z)\simeq 1-\frac{\alpha}{\pi}\exp\left[-\alpha-2n\left(\sinh
  \alpha-\alpha \right)\right] 
\end{equation}
with $\alpha=\textrm{arccosh}\left(1/2z^2-1\right)$.
\begin{figure}[t!]
  \centering
  \epsfig{figure=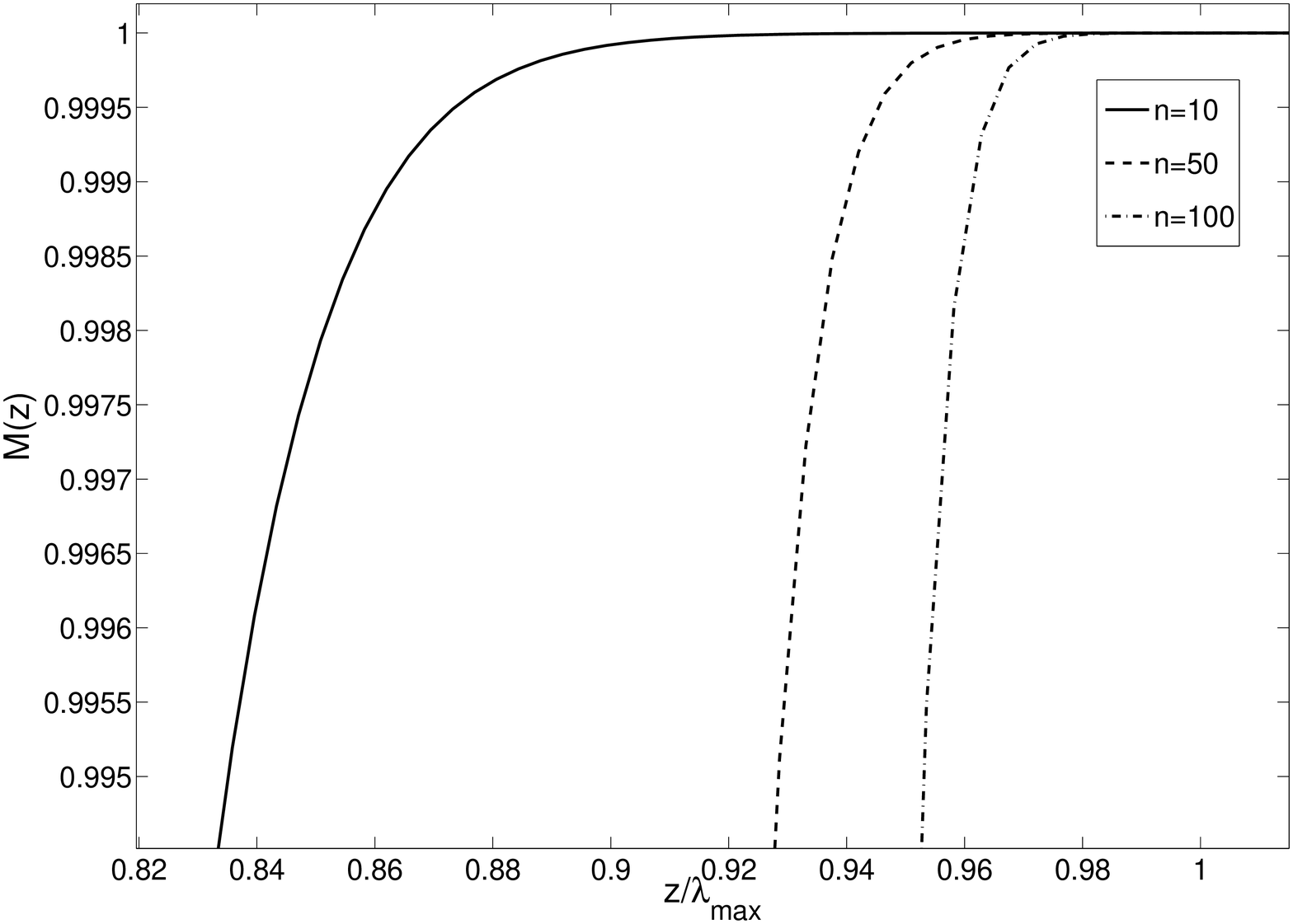,width=0.5\textwidth}
  \caption{Comparison between three analytical solutions for $M(Z)=1$, it is
    possible to see how, for decreasing disorder ($n$ increasing) the solution
    approaches the solution of the problem without randomness. In the $x$-axis
    the variable $z$ has been normalized with respect to the maximum
    eigenvalue $\lambda_{max}$ }
  \label{fig:intD2}
\end{figure}
Fig. \ref{fig:intD2} shows the behavior of the integrated density of states
in the vicinity of the band edge at $\lambda_{\max}$, for three different
values of the disorder strength. In Ref. \cite{A:Buonsante06CM} the behavior
of the corresponding density of states is related to the presence of a BG
phase outside the MI region. In that case a possible direct transition from
the MI to the SF phase is possible for small disorders and specific values of
the chemical potential, and signaled by a singularity at the band edge in the
density of states, similar to the Van Hove singularity characterizing the
homogeneous case, Eq. \eqref{eq:Minf}. Conversely, in the present case, where
density of states depends on the chemical potential through an overall
multiplicative factor, an infinitesimal disorder is sufficient to smear the
discontinuity in the density of states. Hence an intermediate BG phase is
expected for every value of the chemical potential, which is in agreement with
the previously noted shape of the BG phase in Figs. \ref{fig:lobe}and \ref{fig:lobStab}.

\begin{figure}[t!]
  \centering
  \epsfig{figure=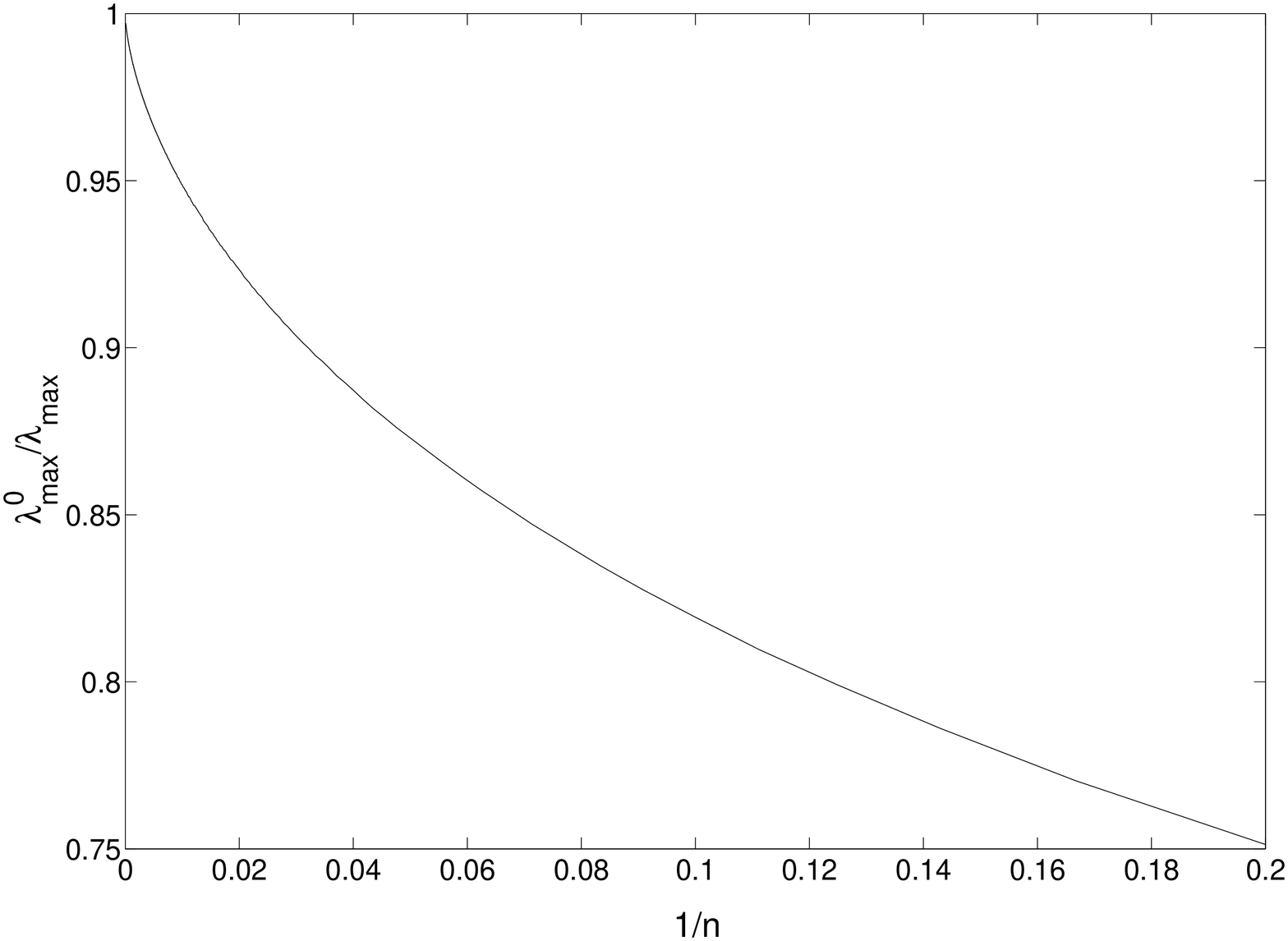,width=0.5\textwidth}
  \caption{Plot of the ``shrinking factor'' for the MI lobe boundary as a
  function of the disorder intensity ($1/n \rightarrow 0$ represents the case
  without disorder).}
  \label{fig:shFact}
\end{figure}

\section{Conclusions}
\label{sec:concl}
In this paper we have considered the effect of a random hopping term on the
zero-temperature phase diagram of the Bose-Hubbard Hamiltonian. Analogously to
what happens when a random on-site term is considered, we have observed the
emergence of a Bose-glass phase. The analysis has been performed within a
mean-field approach both numerically and analytically. The boundaries of the
Mott lobes and the presence of a surrounding BG phase has been related to the
spectral feature of an off-diagonal Anderson model. For the future, we plan to
extend our research towards the finite-temperature case and to higher
dimensions.

{\bf Acknowledgments.} One of the authors (PB) acknowledges a grant form
\textit{Lagrange project}-CRT Foundation and hospitality of the Ultra Cold
Atoms group at the University of Otago.  The work of F.M. has been entirely
supported by the MURST project Cooperative Phenomena in Coherent System of
Condensed Matter and their Realization in Atomic Chip Devices.

\end{document}